# Li-decorated BC$_3$ nanopores: Promising materials for hydrogen storage

I. Cabria[a], A. Lebon[b], M. B. Torres[c,*], L. J. Gallego[d], A. Vega[a]

[a]*Departamento de Física Teórica, Atómica y Óptica, Universidad de Valladolid, ES-47011 Valladolid, Spain*
[b]*Laboratoire de Chimie Electrochimie Moléculaire et Chimie Analytique, Université de Brest, UMR CNRS 6521, F-29285 Brest, France*
[c]*Departamento de Matemáticas y Computación, Escuela Politécnica Superior, Universidad de Burgos, ES-09006 Burgos, Spain*
[d]*Área de Física de la Materia Condensada, Departamento de Física de Partículas, Facultad de Física, Universidad de Santiago de Compostela, ES-15782 Santiago de Compostela, Spain*

**Abstract**

In the quest of new absorbent for hydrogen storage, we investigate the capacities of slit pores formed by two BC$_3$ sheets decorated with Li atoms. Their hydrogen storage capacities are determined using density- functional theory in conjunction with a quantum-thermodynamic model that allows to simulate real operating conditions, i.e., finite temperatures and different loading and depletion pressures applied to the adsorbent in the charge-delivery cycles. We show that the capacities of the adsorbed hydrogen phase of Li-decorated BC$_3$ slit pores are larger than those reported recently for graphene and Li-decorated borophene slit pores. On the other hand, the usable volumetric and gravimetric capacities of Li-decorated BC$_3$ slit pores can meet the targets stipulated by the U.S. Department of Energy (DOE) for onboard hydrogen storage at moderate temperatures and loading pressures well below those used in the tanks employed in current technology. In particular, the usable volumetric capacity for pore widths of about 10 Å meets the DOE target at a loading pressure of 6.6 MPa when depleting at ambient pressure. Our results highlight the important role played by the rotational degree of freedom of the H$_2$ molecule in determining the confining potential within the slip pores and their hydrogen storage capacities.

## 1. Introduction

One of the most important challenges in the XXI century is the energy. The enormous consumption of fossil fuels such as petroleum, natural gas and coal derived from the needs of modern society, and the ecological problems related to the pollution and greenhouse gases, have led to the need to find new low-cost, environmentally-friendly energy sources that can support the sustainability and the global energy demands. In this context, hydrogen has emerged as a promising alternative to replace fossil fuels in the future. Some companies and governmental and non-governmental institutions across the globe are developing strategies and making massive investments toward the "hydrogen economy" [1–4], which is expected to be a socioeconomic reality on a relevant scale in the next decades.

---

*Corresponding author
*Email address:* begonia@ubu.es (M. B. Torres)

*November , 2023*

Hydrogen is really an ideal fuel because it is lightweight (it has the highest energy density of all common fuels by weight, but the lowest energy density by volume), highly abundant, and the by-product of its combustion is simply water vapour (i.e., no $CO_2$ and other harmful gases are produced). The overwhelming majority of hydrogen on earth is chemically bound as $H_2O$. Thus, whether hydrogen can be considered a clean form of energy or not depends on the primary energy that is used to split water. A way to obtain "green hydrogen" is to use wind or sunlight in combination with photovoltaic cells to produce water-splitting electrolysis [5–7]. However, the use of hydrogen energy at a large scale still needs some scientific and technological barriers to be overcome. The most critical problem is storage, in spite of the advances that have already been made in this area during the last decades [2, 8–13].

High-pressure storage of hydrogen in the gas form, cryogenic storage in liquid form and hydride-based storage in the solid form have several inconveniences [3, 14, 15], and in the last years other possibilities have been explored. In particular, extensive theoretical efforts have been performed to investigate the hydrogen storage ability of carbon nanostructures, which are light and have a high accessible surface area, such as carbon nanotubes [16, 17], graphene [18], pillared graphene [19], sandwiched graphene-fullerene composites [20] and graphene slit pores [21]. These last nanostructures can be considered as simple models for the pores existing in nanoporous carbon materials like activated carbons [22] and carbide-derived carbons [23]. Experimental studies have been performed on several carbon nanostructures such as single-walled carbon nanotubes [24], nanoporous spongy graphene [25], few-layer graphene-like flakes [26] and nanoporous activated carbon cloths [27], among others. In microporous frameworks, by optimizing the pore size of the carbon nanostructures, it is possible to increase the hydrogen affinity due to the overlap of the potentials from the pore walls.

There are specific requirements for a material to be a good candidate for onboard hydrogen storage. The targets stipulated by the U.S. Department of Energy (DOE) for 2025 are a reversible gravimetric hydrogen storage capacity of 5.5 wt.% and a reversible volumetric hydrogen storage capacity of 0.040 kg/L at room temperature and moderate pressure [28]. To achieve reversible storage of $H_2$ at those conditions, the binding energy needs to be 0.2–0.6 eV per $H_2$ molecule, i.e., in a range between physisorption and chemisorption [29, 30]. We note that the DOE 2025 targets refer to the usable capacity, also known as the working or delivery capacity. The usable capacity is defined as the difference between the capacity of the full tank at maximum pressure (the loading pressure) and the capacity remaining in the tank at the depletion, back or minimum pressure required to run the fuel cell (the depletion pressure) [28, 31–33]. Today's fuel cell cars use compressed hydrogen gas, squeezing about 5 kg into a 70 MPa carbon fibre reinforced tank. Such high loading pressure requires to use a heavy and expensive fuel container, besides a lot of energy to compress the gas. The goal, therefore, is to find an appropriate adsorbent with a good capacity at a lower loading pressure, allowing to use light-weight, inexpensive, and more conformable pressure tanks as well as to reduce the energy penalty for compressing the gas. The depletion pressure of most hydrogen systems typically lies between 0.3 and 0.8 MPa, but it could be reduced to ambient pressure.

In general, the studies performed so far on carbon nanotubes and graphene-based systems have shown that those materials, in pristine form, do not fulfill the $H_2$-binding energy criterion. However, it has been proposed that those nanostructures could be used to construct hydrogen storage media if they are decorated with atoms of carefully selected species, such as early transition, alkali and alkali-earth metals (see, e.g., Refs. [34, 35] and those cited therein). It should be pointed out that those theoretical studies give predictions for the total and/or the excess capacities, not for the usable or delivery capacities.

Concurrently with the research on carbon-based nanostructures, there has been a growing interest in recent years in two-dimensional (2D) materials beyond graphene, and some of them have been investigated for their possible use as hydrogen storage media, including phosphorene [36–39], borophene [40–42] (two examples of monoelemental 2D materials with different properties from those of graphene), carbon nitrides[29, 30] and MXnes [43, 44], among others. The case of borophene is particularly striking: boron is carbon's neighbour in the periodic table and has similar valence orbitals, but its electron deficiency prevents it from forming a graphene-like honeycomb structure. However, due to the similar atomic radius of boron and carbon (and to the fact that boron atoms can substitute carbon atoms in the graphene structure without activation barrier [45]), it is possible to form substitutional boron doped ordered forms of graphene without altering practically the honeycomb geometry. In fact, recent phonon spectra and ab initio molecular dynamics calculations have shown that a graphene-like $BC_3$ sheet composed of hexagonal carbon rings connected to its six neighboring carbon rings through six isolated boron atoms is mechanically and thermally stable in free standing form [46]. However, it should be pointed out that the experimental realization of the $BC_3$ sheet dates back to 1986, when it was synthesized for the first time as a lustrous film of metallic appearance
by the chemical reaction of benzene and boron trichloride at 800 °C [47]. Subsequently, a $BC_3$ honeycomb sheet with excellent crystalline quality was grown over the entire macroscopic surface area of $NbB_2(0001)$ [48–51]. Strategies to keep the structural integrity, i.e., the absence of defects, in the synthesis of the $BC_3$ monolayer to be used in practical applications (for instance, as gas sensor [52]) have recently been proposed by Yong et al. on the basis of first-principles calculations [53].

Unlike graphene, in which the $\pi$ electrons are located over the entire carbon lattice, the carbon rings of the $BC_3$ sheet have an aromatic character, with six $\pi$ electrons located over each carbon atom, and they are separated by anti-aromatic hexagons composed by boron and carbon atoms without $\pi$ electrons. The electron deficiency of each boron atom creates a "hole-carrier" in the valence band of the bulk $BC_3$ material, thus producing greater electric conductivity than that of graphite [47]. The $BC_3$ sheet has a semiconducting behavior, and the observed metallic appearance of the bulk sample arises from the interaction between neighboring $BC_3$ layers [54].

The hydrogen storage capacity of the Li-decorated $BC_3$ sheet has been studied recently by means of standard density-functional theory (DFT) calculations [55–58]. Those studies reveal that the Li atoms are strongly adsorbed on the $BC_3$ sheet without clustering, which facilitates the reversible hydrogen adsorption and desorption. More specifically, in Ref. [58] it was shown that the binding energy of the Li atom on the $BC_3$ sheet exceeds the cohesive energy of the Li bcc bulk structure. This is because the boron atom has one electron less than the carbon atom, so that the $BC_3$ sheet is an electron-deficient system, making the Li atom to be strongly bounded. Additionally, specific calculations showed that when two Li atoms are adsorbed on the $BC_3$ sheet, the configuration in which the Li atoms stay together is 0.35 eV higher in energy than that when they remain isolated, indicating that the Li atoms can be individually dispersed on the $BC_3$ sheet [58]. On the other hand, the results of Ref. [58] show that not only the decorating metal atom but also the $BC_3$ sheet plays an important role in the hydrogen storage process, because the boron atoms of the sheet improve the induced electric field between the adatoms and the sheet. It should be pointed out that the DFT studies performed in Refs. [55–58] provide information on whether or not the binding energy enters the window required for reversible adsorption as well as the total gravimetric hydrogen storage capacity (not the usable capacity). However, the DFT is a formalism for the ($T = 0$ K) ground state, whereas the experiments and the sorption-desorption process carried out for porous materials, which are the common adsorbents, take place at finite temperature and moderate pressures.

The aim of the work presented here is to provide realistic predictions for the volumetric and gravimetric hydrogen storage capacities of a porous material containing Li-decorated $BC_3$ slit pores. To incorporate finite temperature and pressure, we combine results obtained from DFT calculations with those inferred from a quantum-thermodynamic model [21, 59–62]. A similar methodology has been used recently by us to investigate the hydrogen storage capacity of Li-decorated borophene slit pores, which were found to be optimal for hydrogen storage at low temperature [59]. As it will be shown, an absorbent material containing Li-decorated $BC_3$ slit pores would allow to use loading pressures well below 70 MPa. But, in a wider context, what may be even more relevant is that the methodology employed in this paper can be used to give accurate predictions of the usable capacities of porous materials in general, thus providing a useful guide for the design of efficient systems for hydrogen storage.

Regarding the way we use to simulate the local environment of a porous material based on $BC_3$, it is important to remark that the pioneering experiments by Franklin [63] and the results of more recent experimental studies on nanoporous carbons [64, 65] showed that these materials are mainly composed by slit-shaped pores, with a pore width distribution in the range of several to hundred angstroms. Previous calculations for graphene slit pores (i.e., two parallel graphene sheets separated by a distance in the range of nanometers) confirm that the local environment of porous carbon materials is well captured by means of such slit pores [21, 60, 61]. In the present work, slit pores made of two parallel $BC_3$ sheets in their most stable honeycomb configuration are used in the simulations. As indicated above, the stability of this sheet has recently been confirmed by phonon spectra and ab initio molecular dynamics simulations [46], and it has also been tested by us as a preliminary step of our calculations. The main objective of the present work is to find the optimal values of the pore width, temperature and pressure for efficient volumetric and gravimetric hydrogen storage capacities of an hypothetical $BC_3$-based porous material. To the best of our knowledge, no experimental data on these kinds of systems are yet available, and we expect that our predictions can stimulate the investigation of their hydrogen storage capacities in the laboratory.

The remaining of the paper is organized as follows. Section 2 is divided in three subsections, where we give some details of the DFT approach used, of the quantum-thermodynamic model employed, and of the method applied to compute the adsorbed, total and usable hydrogen storage capacities. In Section 3 we describe the structural results and the method used to determine the confining potential of the $H_2$ molecule in the Li-decorated $BC_3$ slit pore. In Section 4 we present and discuss our results for the adsorbed and total hydrogen storage capacities of Li-decorated $BC_3$ slit pores. Section 5 is devoted to the presentation and discussion of the usable hydrogen storage capacities of these kinds of pores. Finally, in Section 6, we summarize the main conclusions of our work.

## 2. Computational methodology

### 2.1. Technical details of the DFT calculations

To optimize the structures, as well as to calculate the three-dimensional interaction potential energy between a $H_2$ molecule and one layer of the Li-decorated $BC_3$ slit pore, $V(x, y, z)$, we performed DFT calculations as implemented in the VASP code [66, 67], which solves the Kohn-Sham equations within the projector-augmented wave (PAW) approach [68]. A cut-off energy of 500 eV was used for the plane-wave basis and the Brillouin zone was sampled using a $6 \times 6 \times 1$ $k$-point grid after the corresponding k-sampling convergence test. The width of the Gaussian smearing was 0.01 eV. A distance of 30.308 Å between periodic images of the sheet along the normal to the surface ensures that they do not interact. The GGA(PBE)+D3 functional [69] was used in the calculations, which includes dispersion interaction corrections. These corrections are essential for describing weakly interacting systems. Relaxations were performed using the conjugate-gradient method until the remaining force acting on the atoms was less than 0.01 eV/Å. The accuracy for convergence of total energies was $10^{-5}$ eV/unit cell. Dipole correction of the total energy was used to remove spurious dipole interactions between the periodic images.

Before determining the potential confining the $H_2$ molecule in the Li-decorated $BC_3$ slit pore, we calculated the stable configuration of the pristine $BC_3$ sheet using the DFT methodology described above. Subsequently, we investigated the optimal Li decoration of the $BC_3$ sheet, and, finally, we performed fully relaxed calculations to determine the most stable locations of a $H_2$ molecule on the Li-decorated $BC_3$ sheet. The results of all these calculations are presented and discussed in Section 3.

*2.2. Quantum-thermodynamic model*

The quantum-thermodynamic model used in this work [21, 59–61], which is an improved version of that proposed by Patchkovskii et al. [62], is based on the thermodynamic equilibrium between the adsorbed and the compressed phases of the hydrogen gas inside the slit pore. That is, the $H_2$ molecules stored by physisorption on the pore surfaces form the adsorbed phase, and the $H_2$ molecules stored only by compression (without interacting with the pore surfaces) form the compressed phase.

Fig. 1 is a schematic representation of a slit pore formed by two parallel Li-decorated $BC_3$ sheets. Through this paper, we shall consider that the width of the Li-decorated $BC_3$ pore is the distance $w$ between the Li layers. The volume of the slit pore is $V_{pore} = S\,w$, where S is the surface of the sheet. The phases of hydrogen stored inside the pore are shown at the right-hand side of the same figure. The (pore-width dependent) full interaction potential of the $H_2$ molecule inside the slit pore is the sum of the interaction potentials of the $H_2$ molecule with the two parallel layers, placed from each other at a distance $w$, i.e., $V_{slit\ pore}(x, y, z; w) = V(x, y, z) + V(x, y, w - z)$, where the two terms are calculated as described in Section 3.

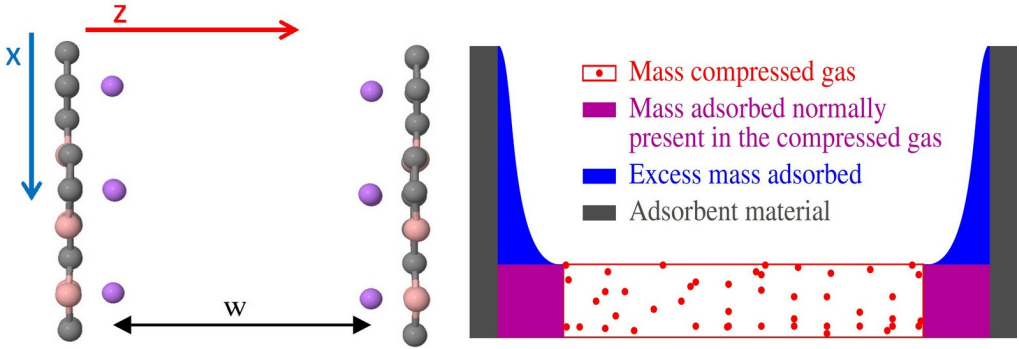

**Fig. 1.** Li-decorated $BC_3$ slit pore of width $w$. B, C and Li atoms are shown with sepia, gray and purple balls, respectively. The panel on the right is a schematic picture of the masses involved in a slit pore.

It is worth to note that the mass of hydrogen stored in the slit pore (the total mass of the stored hydrogen) is the sum of the masses of the adsorbed hydrogen (sum of blue and violet masses in Fig. 1) and the compressed hydrogen (red molecules in Fig. 1), not included in the adsorbed phase. The blue, violet and red masses of Fig. 1 correspond respectively to those of the A, B and C regions shown in the lower panel of Fig. 1 of the work by Parilla et al. [70]. Hence, the mass of the "adsorbed hydrogen" of our work corresponds to the "absolute adsorption" (A+B in the work by Parilla et al.) and the "mass compressed" gas corresponds to the mass of the "free gas" (C in that work). The adsorbed and compressed hydrogen phases occupy different volumes, $V_{adsorbed}$ and $V_{compressed}$, respectively.

Since the slit pore creates a confining region of nanometric dimensions for the $H_2$ molecules, quantum confinement effects arise, i.e., the energy of the $H_2$ molecule is quantized. The energies $E_i$ of the quantum states of the hydrogen molecule are calculated by solving the corresponding Schrödinger equation. The partition function of the adsorbed hydrogen phase, $Z_{ads}$, is then obtained as

$$Z_{ads} = \sum \exp\left(-\epsilon_i/k_B T\right) \quad (1)$$

where $k_B$ is the Boltzmann constant. The equilibrium constant between the adsorbed and compressed phases is given by

$$K_{eq} = \frac{Z_{ads}}{Z_{com}} \qquad (2)$$

where the partition function of the compressed (or free) phase, $Z_{com}$, is obtained as

$$Z_{com} = (d - 2d_{excl})\sqrt{2mk_B T/h^2} \qquad (3)$$

being $m$ the mass of one hydrogen molecule and $d_{excl}$ an exclusion distance due to the steep repulsive part of the interaction potential near the pore layers. In the thermodynamic equilibrium, the equilibrium constant $K_{eq}$ is related to the pressures of the compressed (or external) and adsorbed phases, $P_{ext}$ and $P_{ads}$, through the equation

$$\ln K_{eq} = 1/RT \int v_{mol}(p,T) dP \qquad (4)$$

where $R$ is the gas constant and $v_{mol}(P, T)$ is the molar volume of the compressed hydrogen, which is given by the Mills-Younglove equation of state of $H_2$ [21, 61, 71, 72]

$$v_{mol}(P,T) = \sum c_i f_i(P,T) \qquad (5)$$

where $f_i(P, T) = g_j(P)h_k(T)$, with $j = int((i − 1)/25)$, $k = i − l$ and $l = 25 j$. There are 30 $g_j(P)$ functions
and 25 $h_k(T)$ functions, with $j$ running from 0 to 29 and $k$ from 1 to 25. The units are: $v_{mol}(P, T)$ in $cm^3/mol$, $P$ en MPa and $T$ in Kelvin. The coefficients $c_i$ and the functions $g_j(P)$ and $h_k(T)$ can be found in the Supporting Information of Ref. [21]. By solving Eq. 5, we obtain $P_{ads}$ for each value of $P_{ext}$, and from the Mills-Younglove equation of state we find the molar volume of the adsorbed hydrogen phase, $v_{mol}(P_{ads}, T)$. The masses of the two phases are calculated when the thermodymanic equilibrium between them is reached. We note that the use of a real equation of state for $H_2$ (the Mills-Younglove equation of state contains 750 parameters and enables accurate description of the states of $H_2$ in wide ranges of temperature and pressure [21]) allows to introduce, in an indirect way, the interactions between the $H_2$ molecules in the free or compressed phase. A comparative analysis of the gravimetric and volumetric hydrogen storage capacities of nanoporous carbons, simulated as graphene slit-shaped pores, has recently been made in Ref. [73] using various theoretical models.

In the present work, the quantum-thermodynamic model has been applied to calculate the volumetric and gravimetric hydrogen storage capacities of Li-decorated $BC_3$ slit pores of widths between 4.6 and 40 Å, temperatures between 77 and 350 K, and pressures between 0.5 and 70 MPa.

*2.3. Definitions of the volumetric and gravimetric hydrogen storage capacities*

The volumetric capacity, $v_c$, of the adsorbed hydrogen phase at $P$ and $T$ is calculated, in kg of $H_2$/L, by means of the equation

$$v_C = v_C(P, T) = \frac{mass_{H\,adsorbed}(P,T)}{V_{pore}} \qquad (6)$$

being $mass_{H\,adsorbed}(P, T)$ the mass of the adsorbed hydrogen in kg and $V_{pore}$ the volume of the pore in liters. The mass of the adsorbed hydrogen is obtained, in kg, as

$$mass_{H\,adsorbed} = \frac{M(H_2)}{v_{mol}(P_{ads},T)} V_{adsorbed} \qquad (7)$$

where $V_{adsorbed}$ is the volume, in liters, of the adsorbed phase, $M(H_2)$ is the molar mass of $H_2$ in kg/mol, 0.00201588 kg/mol, and $v_{mol}(P_{ads}, T)$ is the molar volume of the adsorbed phase in L/mol. $V_{adsorbed} = S\, w_{ads}$, where $w_{ads}$ is the width of the adsorbed phase, computed as indicated in detail in Ref. [59].

The gravimetric capacity, $g_c$, of the adsorbed hydrogen phase at $P$ and $T$ is calculated, in $wt.\%$, as

$$g_c = g_c(P, T) = \frac{100\, mass_{H\,adsorbed}(P, T)}{mass_{H\,adsorbed}(P,T) + mass_{adsorbent\,material}} \qquad (8)$$

where $mass_{adsorbent\,material}$ is the mass of the adsorbent material (i.e., that of the slit-shaped pore) in kg. The slit pore is composed by *two sheets* and this is taken into account in the calculation of $mass_{adsorbent\,material}$ and $g_c$.

The total volumetric and gravimetric capacities are given by similar equations to those of the adsorbed hydrogen phase, but using the total mass of the stored hydrogen, $mass_{H\,total}(P, T)$, i.e.,

$$v_{c,total} = v_{c,total}(P, T) = \frac{mass_{H\,total}(P, T)}{V_{pore}} \qquad (9)$$

$$g_{c,total} = g_{c,total}(P, T) = \frac{100\, mass_{H\,total}(P, T)}{mass_{H\,total}(P, T) + mass_{adsorbent\,material}} \qquad (10)$$

where $mass_{H\,total}(P, T)$ includes the mass of the adsorbed hydrogen phase and that of the compressed hydrogen phase, not included in the adsorbed phase. The latter is obtained as the product of the volume occupied by the compressed hydrogen, $V_{compressed}$, and the free-hydrogen density.

The capacities defined above give the amount of total hydrogen that can be stored at a given pressure and temperature. For practical purposes, however, the relevant quantities are the usable capacities, which are those to which the DOE 2025 conditions refer. The usable mass of hydrogen stored at a given pressure $P$ (loaded state) and temperature $T$ is the difference between the total mass of hydrogen stored at these $P$ and $T$ and the total stored mass at the depletion pressure and the same temperature $T$ [28, 31–33]. The usable volumetric and gravimetric capacities are defined using the usable mass of stored hydrogen.

### 3. Structural results and determination of the confining potential

Before explaining the way we used to determine the confining potential, we discuss the predicted structures of the pristine and the Li-decorated $BC_3$ sheets. The pristine $BC_3$ sheet is constituted by boron hexagons with carbon hexagons inside (left-hand panel of Fig. 2). The primitive cell of the $BC_3$ sheet has eight non-equivalent atoms, of which six are carbon atoms and two are boron atoms. The upper panel of Fig. 2 shows the supercell structure used in our calculations (with interatomic distances obtained after relaxation), which has three times the size of the $BC_3$ primitive cell. We found that the optimal Li-decorated structure is similar to that reported by Zhao et al. [56] using the LDA(CA) functional [74, 75], i.e., it has four Li atoms adsorbed on one side of the supercell, occupying the central positions of the external rings (right-hand panel of Fig. 2). The distance between the Li atoms and the $C_6$ hexagon was found to be 1.57 Å, a value that increased until 1.63 Å for the $C_4B_2$ hexagon and the Li-Li distance is 4.48 Å. The average binding energy of the Li atoms on the sheet, calculated as $E_b = (E(BC_3) + 4E(Li) - E(Li\text{-}BC_3))/4$, is 1.81 eV, which is larger than the experimental cohesive energy of bulk Li (1.63 eV [76]). We note that the binding energy of a Li atom on graphene has been reported to be 1.096 eV [77].

As the boron atom has one less electron than the carbon atom, the $BC_3$ sheet is electron deficient, so that the Li atoms are strongly bonded to the $BC_3$ sheet due to an electronic transfer from them to the sheet.

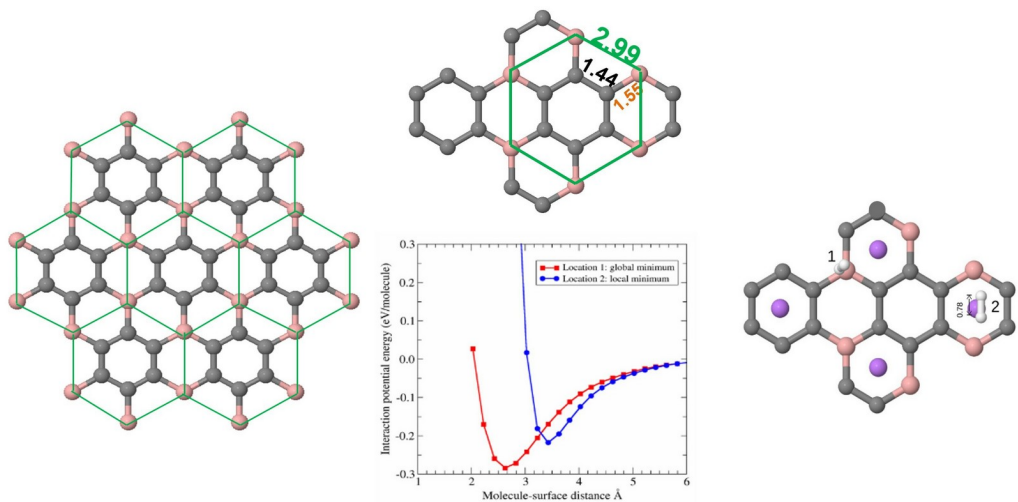

**Fig. 2.** Structures of the $BC_3$ sheet (lef panel) and of the supercell used in our calculations after relaxation (upper panel). The right panel shows the (supercell) structure of the Li-decorated $BC_3$ sheet, with $H_2$ molecules at the locations corresponding to the global minimum (1) and a local minimum (2) of the interaction potential between the $H_2$ molecule and the Li-decorated $BC_3$ sheet. The lower panel shows the interaction potential curves around those minima. B, C, Li and H atoms are shown with sepia, gray, purple and white balls, respectively, and the numbers are atom-atom distances in Å.

We performed a series of tests to confirm the dynamic stability of the Li-decorated $BC_3$ structure, i.e., the absence of Li clustering effects. This is an important point to be taken into account when analyzing the ability of a metal-decorated material for hydrogen storage. To this end, we first computed several diffusion paths for the movement of a Li atom of the Li/$BC_3$ system on the $BC_3$ surface, and the results are shown in Fig. 3. It can be seen that the diffusion of the Li atom involves energy barriers of more than 0.5 eV. These energy barriers are significant and sufficient to block the movement of the Li atoms around their equilibrium positions, since room temperature corresponds to about 0.025 eV. A similar effect has been observed in the case of the Li-popgraphene system [78].

On the other hand, we investigated possible configurations of the Li/$BC_3$ system with a $Li_2$ dimer and a $Li_4$ cluster, with rhombic and tetrahedral geometries, to check their relative stabilities. Fig. 4 shows the relaxed structures and their energies with respect to that of the structure shown in the right-panel of Fig. 2 (again displayed in Fig. 4 as structure A). The structure with a $Li_2$ dimer placed on the $BC_3$ sheet (structure B) is energetically 0.44 eV less favorable than the structure A. The structure in which the four Li atoms are placed above the central hexagons of the $BC_3$ sheet forming a rhombus (C) and the tetrahedral structure (D) are respectively 0.75 eV and 1.62 eV less favorable than the structure A. It should also be pointed out that if a free $Li_2$ dimer is placed above the $BC_3$ sheet and it is allowed to relax, the dimer dissociates when it approaches the sheet surface without energy barrier.

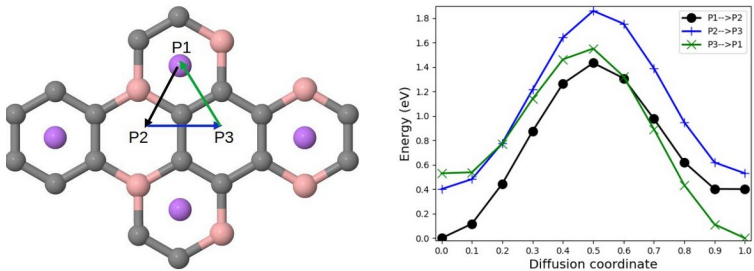

**Fig. 3. Diffusion paths for the movement of a Li atom of the Li/BC$_3$ system on the sheet surface, starting from its equilibrium position, and their corresponding energy profiles (right). B, C and Li atoms are shown with sepia, gray and purple balls, respectively.**

This occurs independently of the initial orientation of the free Li$_2$ dimer. Similar results are obtained when a Li$_4$ cluster is placed on the BC$_3$ sheet and it is allowed to relax approaching the BC$_3$ surface. The results of all these calculation confirm that the Li atoms tend to be dissociated on the BC$_3$ sheet, and that the most stable structure of the Li/BC$_3$ system is that shown in the right-hand panel of Fig. 2, in keeping with the results reported in Refs. [55–58]

Next, we determined the most stable locations (positions and orientations) of a H$_2$ molecule on the Li-decorated BC$_3$ sheet. Those locations correspond to different local minima of the interaction potential energy, defined as

$$V(x, y, z) = E(H_2@Li\text{-}BC_3) + E(H_2) - E(Li\text{-}BC_3) \tag{11}$$

where $E(H_2@Li\text{-}BC_3)$, $E(H_2)$ and $E(Li\text{-}BC_3)$ are the energy of H$_2$ at $(x, y, z)$ on the Li-decorated BC$_3$ sheet, the energy of an isolated H$_2$ molecule and the energy of the Li-decorated BC$_3$ sheet, respectively, and $z$ is the distance between the H$_2$ molecule and the surface of the sheet. For the interest of the readers, we show in Figs. S1 and S2 of the Supporting Information the locations of the H$_2$ molecule on the Li-decorated BC$_3$ sheet for the first six local minima of the interaction potential, and the interaction potential curves around those minima, as obtained in our calculations using the GGA(PBE)+D3 functional [69] (note that the values given in Fig. S2 are the H$_2$-binding energies, i.e., the sign-reversed potential energies). The global minimum of the potential is at -0.28 eV, and the H$_2$ molecule is situated almost perpendicularly to the Li-decorated BC$_3$ sheet on a carbon atom (location 1 in Fig. 2). Another minimum is at -0.22 eV, and the H$_2$ molecule is on top of a Li atom, almost parallel to the sheet (location 2 in Fig. 2). The H-H distance of the adsorbed H$_2$ molecule is 0.78 Å, a value slightly higher than the experimental value of 0.741 Å [79, 80]) in vacuum, which reveals the interaction between the molecule and the Li-decorated BC$_3$ sheet.

We note that the optimal orientation of the H$_2$ molecule at the different local minima of the interaction potential differs, with a noticeable energy cost when rotating the H$_2$ molecule away from its equilibrium orientation, an important fact to consider for accurate determination of the confining potential, as it will be shown below. For instance, in the local minimum in which the H$_2$ molecule is above a Li atom and its orientation is parallel to the Li-decorated BC$_3$ sheet, the energy penalty when imposing a perpendicular

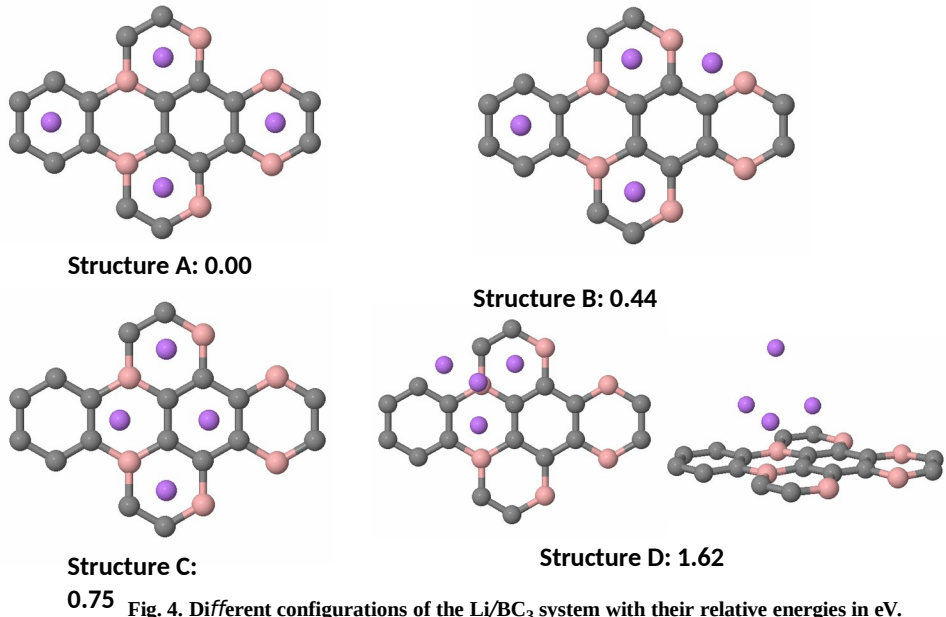

**Structure A: 0.00**

**Structure B: 0.44**

**Structure C: 0.75**

**Structure D: 1.62**

Fig. 4. Different configurations of the Li/BC$_3$ system with their relative energies in eV.

orientation is 0.17 eV, a huge amount considering that for achieving reversible storage the H$_2$-binding energy needs to be 0.2-0.6 eV per molecule. And what is more important, imposing a fixed orientation of the H$_2$ molecule on the whole Li-decorated BC$_3$ sheet prevents it from locating the actual minima of the confining potential. This point will be discussed in detail below.

For comparative purposes, in Fig. S2 of the Supporting Information we show the interaction potential curves around the six main local minima of the potential, as obtained using the LDA(CA) [74, 75], GGA(PBE) [81, 82] and GGA(PBE)+D3 functionals [69]. The dispersion corrections are not included in the LDA(CA) and GGA(PBE) functionals. Our results show that, around the equilibrium positions, the values predicted with the GGA(PBE)+D3 functional are lower than those derived using the LDA(CA) and higher than those obtained with the GGA(PBE). However, for H$_2$-sheet distances larger than about 3.5 Å, a crossing takes place and the GGA(PBE)+D3 functional predicts lower values of the interaction potential.

We now focus on the determination of the confining potential. In order to see the effect of considering a fixed orientation of the H$_2$ molecule, we selected the orientation corresponding to the global minimum of the potential and displaced the molecule, with this orientation fixed, on a *xy* grid, with $\Delta x = \Delta y = 0.20$ Å. Hence, for the *xy* plane, 1081 points were considered in a single point calculation mode. The process was repeated at different values of *z*, from 1.5 to 6.5 Å, with an interval of $\Delta z = 0.20$ Å. This calculation mode turned out to be problematic as the H atom with the lower z could sit very close to one of the Li decorating atoms. This resulted in strongly anti-bonding energy that could even reach tens of eV. The upper panel of Fig. 5 shows the resulting potential for $z = 2.831$ Å.

It is then clear that one has to perform most realistic (and computationally much more demanding) calculations by including the rotational degree of freedom of the adsorbed H$_2$ molecule. To this end, we performed calculations by fixing only the center of the H$_2$ molecule according to its position in the *xy* plane. To be more precise, the height of the center of mass of the H$_2$ molecule with respect to the BC$_3$ layer was selected to be $z = 2.831$ Å, which corresponds to the global minimum of the interaction potential, and the orientation was relaxed for each of the 1081 points of the $xy$ grid, all at $z = 2.831$ Å. In our protocol, the H-H distance was kept fixed to the distance found for the global minimum, that is, $d_{H-H} = 0.78$ Å. The optimal spherical polar and azimuthal angles at each point were determined using the Nelder-Mead simplex algorithm. The lower panel of Fig. 5 shows the confining potential for $z = 2.831$ Å. Not only the actual local minima are now captured, but the interaction potential energy is noticeably more binding. After the computations for the 1081 points at $z = 2.831$ Å, the obtained polar and azimuthal angles for each position on the grid were taken over for the other $z$ values, namely, in the range from $z = 1.5$ to 6.5 Å. Additional calculations were carried out to verify that this is a good approximation. After obtaining the interaction potential energy in the upper half of the cell, the interaction potential energy in the lower half of the cell was obtained by symmetry.

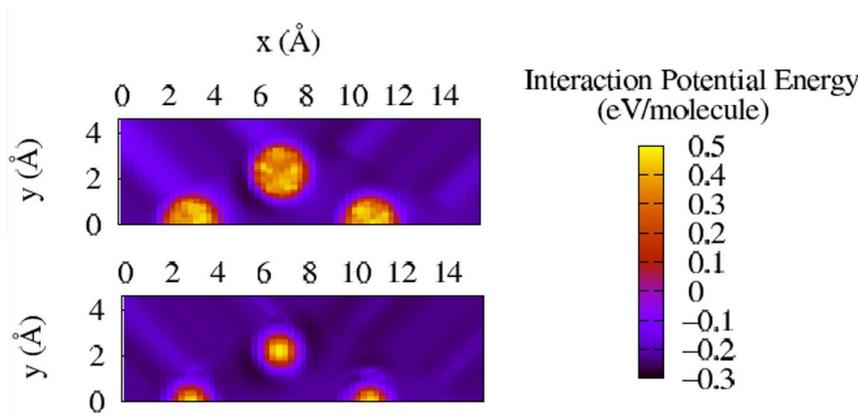

**Fig. 5.** Interaction potential energy between a single $H_2$ molecule in the plane $z = 2.831$ Å and the Li-decorated $BC_3$ sheet, projected on the $xy$ plane. The upper and lower panels correspond to the cases of fixed and non-fixed orientations of the $H_2$ molecule, respectively.

The values of the interaction potential energy obtained in the DFT calculations at the ($x$, $y$, $z$) grid points were fitted to a piece-wise function $V(x, y, z)$. The grid has 1081 ($x$, $y$) different points or regions. The interaction potential energies corresponding to the point or region $i$, ($x_i$, $y_i$), were fitted to an analytic function of the shape

$$V(x_i, y_i, z) = A(i) e^{-B(i)z} + \sum C(i,j) z^{-2j} \qquad (12)$$

The set of the 1081 local or regional functions of this type form a piece-wise analytic function, which is the interaction potential energy $V(x, y, z)$. This piece-wise function was used in the quantum-thermodynamic calculations to obtain the hydrogen storage capacities.

In Fig. 6 we show the deepest $V(z)$ curve among all the $V(z)$ curves that were obtained when the rotational degree of freedom of the $H_2$ molecule was included in calculating the interaction potential between the molecule and the Li-decorated $BC_3$ sheet. The curve is compared with that found for the interaction between the $H_2$ molecule and graphene using the optB88-vdW functional [59]. The $V(z)$ curve for the Li-decorated $BC_3$ sheet has a much larger depth (-0.28 eV/molecule against -0.08 eV/molecule). This fact anticipates, at a qualitative level, the much larger volumetric hydrogen storage capacities of Li-decorated $BC_3$ slit pores, as it will be seen in the following section.

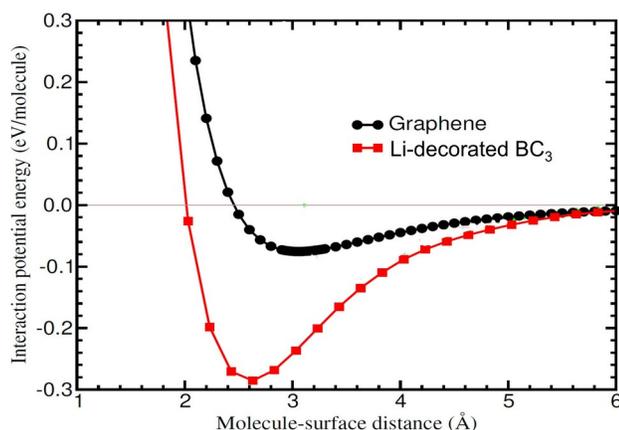

**Fig. 6.** Interaction potential energies between a $H_2$ molecule and graphene and the Li-decorated $BC_3$ sheet as functions of the $H_2$ molecule-surface distance.

## 4. Adsorbed and total hydrogen storage capacities of the Li-decorated BC$_3$ slit pores

The DOE 2025 conditions refer to the usable hydrogen storage capacities, but for a clear understanding of these quantities it is necessary to analyze first the behaviour of the adsorbed and total hydrogen storage capacities and their dependence on the main parameters, pore width, pressure and temperature.

The adsorbed hydrogen storage capacities of Li-decorated $BC_3$ slit pores are shown in Fig. 7 as functions of the pore width, at 298.15 K and pressures of 5 and 25 MPa. For comparison, in the same figure we also show the capacities that we obtained for these kinds of pores when the confining potential was determined by considering a fixed orientation of the $H_2$ molecule, as well as the results that have been reported for graphene slit pores at 298.15 K and 25 MPa [61].

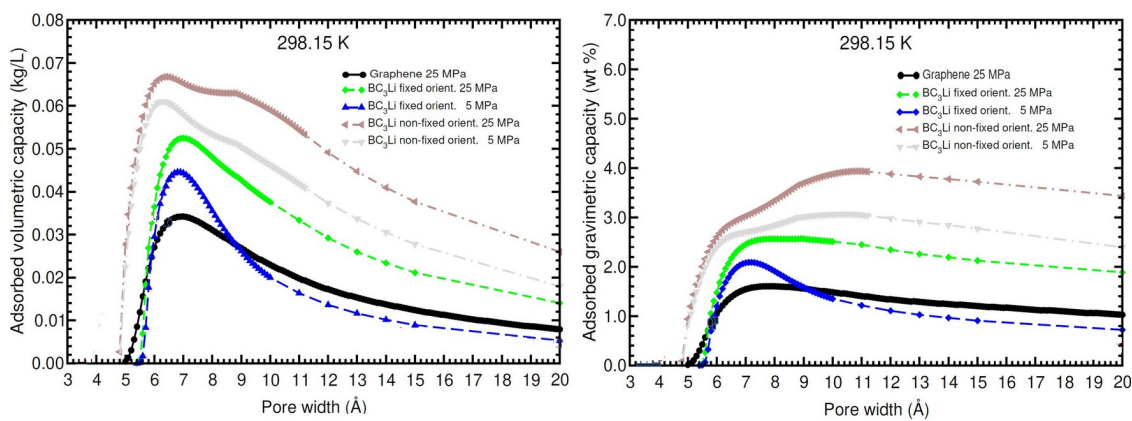

**Fig. 7.** Adsorbed volumetric and gravimetric hydrogen storage capacities of Li-decorated $BC_3$ slit pores as functions of the pore width, at 298.15 K and pressures of 5 and 25 MPa. Also shown are the capacities that have been predicted for graphene slit pores at 298.15 K and 25 MPa [61] (black curves).

We first note that the adsorbed hydrogen storage capacities of Li-decorated $BC_3$ slit pores are much larger than those of graphene slit pores. In fact, at 5 MPa, the former capacities are greater than the latter at 25 MPa.

On the other hand, the adsorbed hydrogen storage capacities of Li-decorated $BC_3$ slit pores that are obtained when the confining potential was derived by imposing a fixed orientation of the $H_2$ molecule are noticeably smaller than those predicted when the molecule is allowed to rotate and accommodate into its most stable orientation. In the latter case, the volumetric capacity has two relative maxima at the two pressures, 5 and 25 MPa, a behaviour not found when the orientation of the $H_2$ molecule is fixed. As indicated in the previous section, the confining potential when the orientational degree of freedom of the $H_2$ molecule is included in the calculations is more binding and captures all the local minima, which causes the volumetric capacity to increase; as the mass of the adsorbed hydrogen increases, the gravimetric capacity increases as well. Fig. 2 shows that, besides the global minimun of the interaction potential between the $H_2$ molecule and the Li-decorated $BC_3$ sheet at -0.28 eV, there is another local minimum at -0.22 eV, in which the $H_2$ molecule is on top of a Li atom. In the slit pore there are two layers of minima of the potential, in correspondence with the two layers of adsorbed Li atoms on each layer. Therefore, there is very likely a correspondence between the topology of the sheet and the evolution of the adsorbed hydrogen storage capacities as a function of the pore width. That peculiarity could be the origin of the shoulders observed in the capacities at a pore width of 8.5 Å (see Fig. 7). Overall, this means that more hydrogen can be stored further away from each $BC_3$ layer of the slit pore, and consequently the capacities increase for wider pores. It should also be noted that the difference between the hydrogen storage capacities of Li-decorated $BC_3$ slit pores at 5 and 25 MPa is rather small for small pore widths, but then increases and reaches a constant value for (relatively) large pore widths, a behaviour that is reflected in the usable hydrogen storage capacities, as it will explained in the next section.

Maximum values of the hydrogen storage capacities at two positions implies that the range of optimal values of the pore width is larger. This latter result is important, because in real nanoporous materials there is not a single pore width but a pore-width distribution, and the experimentalists can be interested in the center and the standard deviation of the distribution. The highest values of the adsorbed hydrogen storage capacities of the Li-decorated $BC_3$ slit pores are summarized in Table 1. Additional information about the adsorbed hydrogen storage capacities can be found in Sects. 2 and 3 of the Supporting Information.

Table 1: Highest adsorbed hydrogen storage capacities of the Li-decorated $BC_3$ slit pores at 298.15 K. $w_{maxv_c}$ and $w_{maxg_c}$ are the corresponding pore widths in Å. The pressure $P$ is in MPa, $v_c$ in kg/L and $g_c$ in wt.%.

| $H_2$ orientation | P | $w_{maxv_c}$ | $v_c$ | $w_{maxg_c}$ | $g_c$ |
|---|---|---|---|---|---|
| non-fixed | 25 | 6.4 | 0.067 | 10.9 | 3.938 |
| non-fixed | 5 | 6.3 | 0.061 | 10.4 | 3.071 |
| fixed | 25 | 7.0 | 0.052 | 8.9 | 2.569 |
| fixed | 5 | 6.8 | 0.045 | 7.2 | 2.090 |

We now discuss the dependence of the adsorbed hydrogen storage capacities on the pressure. The results of our calculations for Li-decorated $BC_3$ slit pores of widths 6.4 and 10.9 Å at room temperature are plotted in Fig. 8. Those pore widths correspond to the highest adsorbed volumetric capacity (0.067 kg/L) and the highest adsorbed gravimetric capacity (3.938 wt.%), respectively, at room temperature and 25 MPa (see Fig. 7 and Table 1).

The shapes of the capacities *vs* pressure curves for both pore widths is similar: at low pressures, both capacities increase very fast, even exponentially, and then they increase slowly reaching constant or asymptotic values at high pressures. However, although the shapes of the volumetric capacity *vs* pressure curves for the two pore widths is similar, there are important differences that will be relevant when studying theusable volumetric capacities.

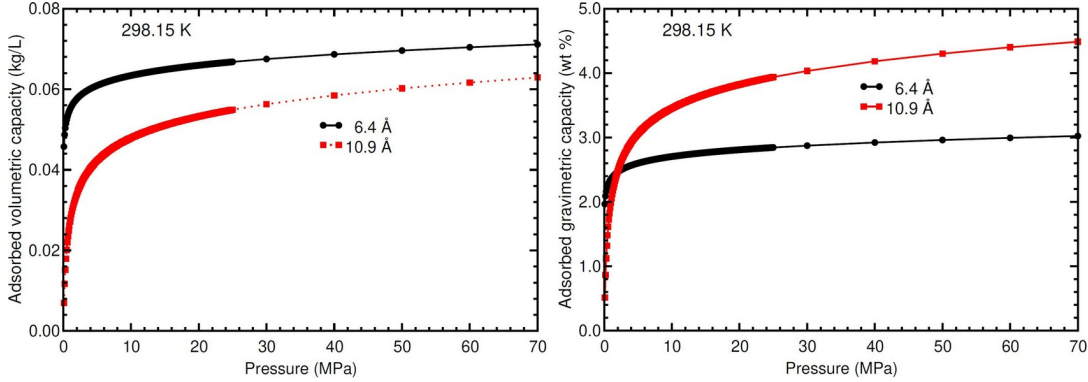

**Fig. 8.** Adsorbed volumetric and gravimetric hydrogen storage capacities of Li-decorated BC$_3$ slit pores of widths 6.4 and 10.9 Å, as functions of the pressure, at 298.15 K.

As it can be noticed in Fig. 8, at low pressures (for instance, 0.5 MPa), the adsorbed volumetric capacity for the 6.4 Å pore is much higher than for the 10.9 Å pore, and hence the difference between the adsorbed volumetric capacities at 25 and 0.5 MPa is much smaller for the 6.4 Å pore than for the 10.9 Å pore. This means that the usable volumetric capacity at 25 MPa for the 6.4 Å pore will be much smaller than that for 10.9 Å pore. Therefore, in spite of the high numerical value of the asymptote of the adsorbed volumetric capacity for the 6.4 Å pore and that this asymptotic value is higher than the asymptotic value for the 10.9 Å pore, the usable capacity is much smaller for the 6.4 Å pore than for 10.9 Å pore. The results of our calculations for the usable hydrogen storage capacities of the Li-decorated BC$_3$ slit pores will be discussed in detail in the next section. It should be noted that, at high pressures, the adsorbed volumetric capacity for the 6.4 Å pore is higher than for the 10.9 Å pore, while the adsorbed gravimetric capacity has the opposite behaviour due to the smaller density of the 10.9 Å pore (larger pore width and, hence, smaller density). More specifically, at high pressures, we obtain $v_c(6.4)/v_c(10.9) = 1.125$, while $g_c(6.4)/g_c(10.9) = 0.667$. Using the approximate Eq. S2 of the Supporting Information, we can write

$$\frac{g_c(6.4)}{g_c(10.9)} \approx \frac{v_c(6.4)}{v_c(10.9)} \frac{6.4}{10.9} \approx 0.661 \quad (13)$$

which is very close to the exact value, 0.667.

As regards the dependence of the adsorbed hydrogen storage capacities on temperature, the results for pore widths of 10 and 35 Å at pressures of 0.5, 25 and 35 MPa and temperatures between 77 and 300 K have been plotted in Fig. 9. Both capacities decrease as the temperature increases for all pore widths and pressures. The capacities are very high at low temperatures, but the important point is the difference between the capacities at $P = 35$ or 25 MPa and at 0.5 MPa. This difference is the main part of the usable capacity. As it can be noticed, at low temperatures the 35-0.5 and 25-0.5 MPa adsorbed capacity differences are much smaller than at room temperatures for the two pore widths studied. For instance, for a pore width of 10 Å the difference is 0.004 kg/L at 77 K, while at 300 K it is 0.036 kg/L, which is about ten times higher. For a pore width of 35 Å, those differences are 0.0016 kg/L and 0.0108 kg/L at 77 and 300 K, respectively. Hence, for the 35 Å pore the adsorbed capacity difference at room temperature is about seven times the difference that exits at low temperature. Another important result is that the adsorbed capacity difference increases as the temperature increases (see Fig. 9). The increase is approximately linear with the temperature for the 10 Å pore and more abrupt for the 35 Å pore. All these results reflect the potential conditions of Li-decorated BC$_3$ nanoporous systems to be used as effective media for hydrogen storage at room temperature.

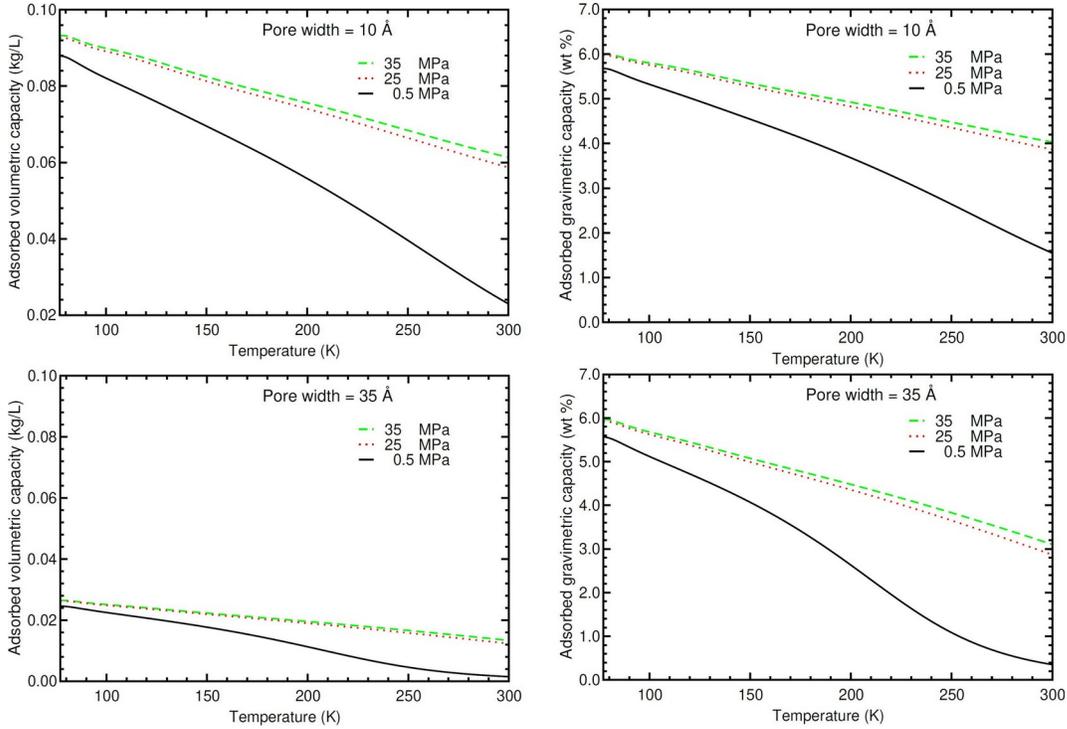

**Fig. 9.** Adsorbed volumetric and gravimetric hydrogen storage capacities of Li-decorated BC$_3$ slit pores of widths 10 and 35 Å, as functions of the temperature, at 0.5, 25 and 35 MPa.

In the remaining of this section, we discuss the dependence of the total hydrogen storage capacities (that is, those that include the contribution of the compressed hydrogen phase) on the pore width, pressure and temperature. We remind that this is the required data for determining the usable hydrogen storage capacities. Some of the main results obtained in our calculations are shown in Fig. 10 by means of colour maps. Additional information, also used in the discussion presented below, is provided in Fig. S5 of the Supporting Information.

The total capacities versus pore width and temperature are shown in the upper panels of Fig. 10 for a pressure of 35 MPa. The highest total volumetric capacities, about 0.1 kg/L, are obtained for pore widths around 11 Å at low temperatures. The highest total gravimetric capacities are about 15 wt.% and are obtained also at low temperatures for larger pore widths, 40 Å. In the upper panels of Fig. S5, we show the results for a lower pressure of 25 MPa. For pore widths between 5 and 15 Å, the volumetric and gravimetric capacities are about 0.06-0.07 kg/L and 4-6 wt.%, respectively. Those values, although very high, should not be compared with the DOE 2025 targets.

The total capacities versus pore width and pressure are shown in the middle panels of Fig. 10 at room temperature, 298.15 K. The highest total volumetric capacity is about 0.06 kg/L, and it is found for pore widths of 5-10 Å and pressures larger than 1 MPa. The highest total gravimetric capacity, about 8 wt.%, is obtained for wide pores, 40 Å, and a high pressure, 70 MPa. At a low temperature of 77 K (middle panels of Fig. S5), the highest total volumetric capacity is about 0.1 kg/L, and it is found for pore widths of 8-12 Å, independently of the pressure. The highest total gravimetric capacity is about 18 wt.% and corresponds

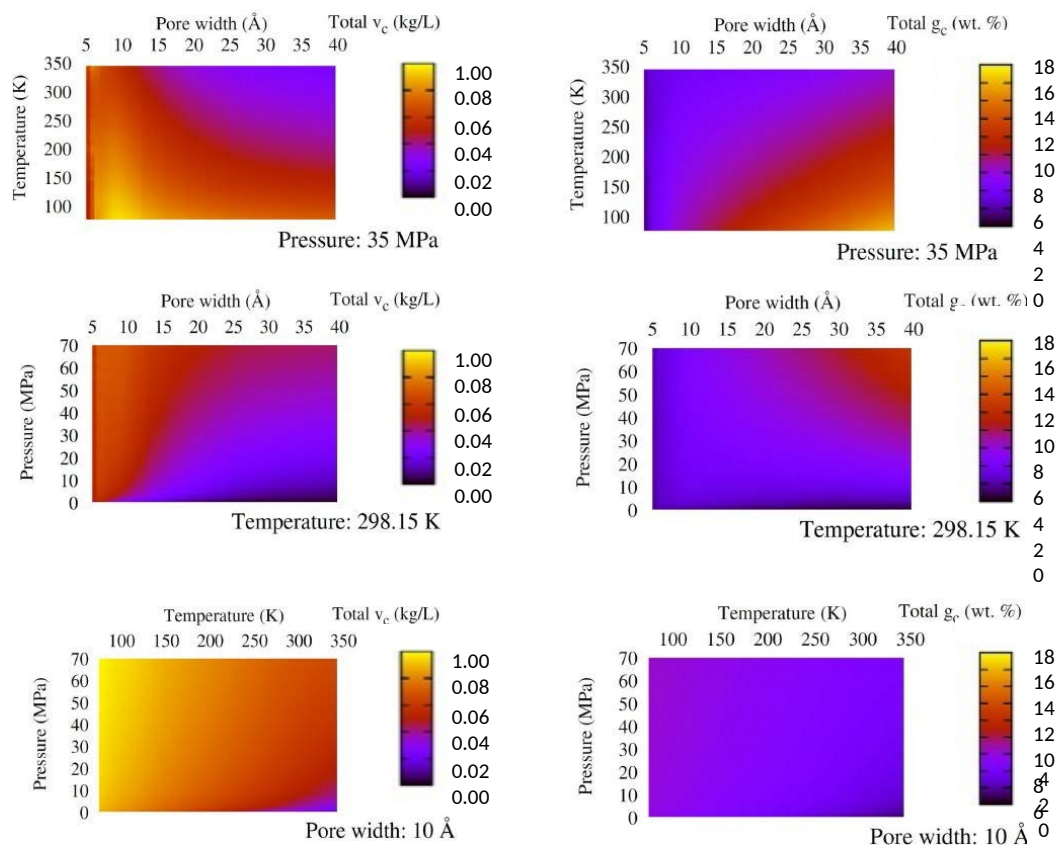

**Fig. 10.** Total volumetric and gravimetric hydrogen storage capacities of Li-decorated BC$_3$ slit pores as functions of the pore width and temperature at 35 MPa (upper panels), of the pore width and pressure at 298.15 K (middle panels) and of the temperature and pressure for a pore width of 10 Å (lower panels).

to wide pores, 40 Å, and pressures equal or larger than 50 MPa.

Finally, the total capacities versus temperature and pressure for a pore of width 10 Å are shown in the lower panels of Fig. 10. The highest total volumetric and gravimetric capacities are about 0.1 kg/L and 7 wt.%, respectively, at 70 MPa and 77 K. The dependence on the pressure and temperature for a pore width of 35 Å (lower panels of Fig. S5) is similar, but the numerical values are different. The highest total volumetric and gravimetric capacities are in this case about 0.06 kg/L and 16 wt.%, respectively, at 70 MPa and 77 K.

## 5. Usable hydrogen storage capacities of Li-decorated BC$_3$ slit pores

The data reported in the previous section were used to calculate the usable hydrogen storage capacities of Li-decorated BC$_3$ slit pores. We compute those capacities as a function of the main variables that could be controlled in the design and operation of these possible nanoporous adsorbents: the pore width, the pressure and the temperature. Our aim is to determine if there are ranges of these variables that satisfy the DOE 2025 requirements. Some results are presented in Fig. 11 by means of colour maps, and additional information is provided in Fig. S6 of the Supporting Information.

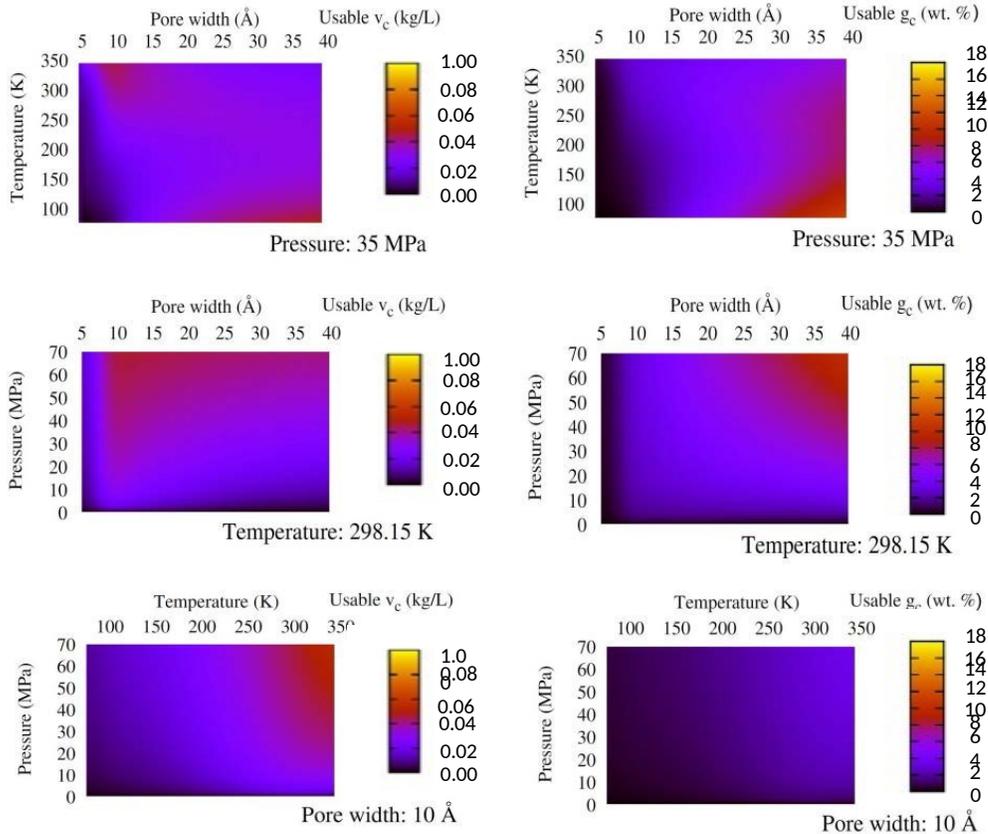

**Fig. 11.** Usable volumetric and gravimetric hydrogen storage capacities of Li-decorated BC$_3$ slit pores as functions of the pore width and temperature at 35 MPa (upper panels), of the pore width and pressure at 298.15 K (middle panels) and of the temperature and pressure for a pore width of 10 Å (lower panels).

The usable capacities have been plotted versus the pore width and the temperature for pressures of 35 MPa (upper panels of Fig. 11) and 25 MPa (upper panels of Fig. S6). The usable volumetric capacity is high and reaches the DOE 2025 target in two different regions at both loading pressures: for narrow slit pores (9-12 Å) at room or near room temperatures (250-350 K), and for wide slit pores (25-40 Å) at low temperatures (77-125 K). On the other hand, for $P$ = 35 MPa, the usable gravimetric capacity reaches the DOE 2025 target in the region of wide pores and the temperature range 77-300 K. Both DOE 2025 targets are reached simultaneously for wide pores (25-40 Å) below room temperature. It is important to notice that, at low temperature, the usable capacities are small for pore widths smaller than 25 Å.

The dependence of the usable capacities on the pore width and pressure can be noticed in the middle panels of Fig. 11 at room temperature, 298.15 K. The usable volumetric capacities reach the DOE 2025 target in a triangular region with $P \geq 50$ MPa and pore widths between 8 and 20 Å, and the usable gravimetric capacities reach the DOE 2025 target in a region of $P \geq 25$ MPa and $w \geq 20$ Å. The usable capacities reach both targets simultaneously at room temperature and high pressures ($P \geq 56$ MPa) for $w \geq 20$ Å. However, it should be noted that the DOE 2025 conditions were established for on board practical applications, i.e., for moderate pressures of about 25-35 MPa or below. It can be seen in the middle panels of Fig. S6 that, at 77 K, the DOE 2025 volumetric target is reached for $P \geq 30$ MPa and $w \geq 25$ Å. The gravimetric target is also reached in the upper right triangular region of the pressure-pore width panel, but in this case for $P \geq 10$ MPa and $w \geq 20$ Å.

Finally, the usable capacities for a pore width of 10 Å are plotted in the lower panels of Fig. 11 versus the thermodynamical variables. The usable volumetric capacity for such a small pore width reaches the DOE 2025 target in the upper right corner of the T-P map, at $T \geq 300$ K and $P \geq 40$ MPa. As regards the usable gravimetric capacity, the results of our calculations indicate that for this small pore width it is not possible to reach the DOE 2025 target in the range of temperatures and pressures considered in our study. For a larger pore width of 35 Å (lower panels of Fig. S6), the usable volumetric capacities reach the DOE 2025 target in the upper left corner, that is, at $P \geq 40$ MPa and $T \leq 200$ K. The usable gravimetric capacities, on the other hand, reach the DOE 2025 target in a large upper region limited by a line from (30 MPa, 77 K) to (40 MPa, 350 K). Both targets are reached simultaneously for this larger pore width at low temperatures and high pressures.

In summary, the analysis presented above indicates that the usable hydrogen storage capacities of Li-decorated $BC_3$ slit pores reach the DOE 2025 targets in several regions of the pore width, temperature and pressure. Temperatures close to room temperature and moderate pressures ($\leq 35$ MPa) define the appropriate operating conditions for practical applications. At those conditions, the usable volumetric capacities of Li-decorated $BC_3$ slit pores reach the DOE 2025 target (0.04 kg/L) for small pore widths (9.0-12.0 Å) and for 30-35 MPa. In those conditions, the usable hydrogen storage gravimetric capacity lies between 2.4 and 3.2 wt. %, which is twice smaller than the DOE 2025 gravimetric target. The usable gravimetric capacities of Li-decorated $BC_3$ slit pore reach the DOE 2025 target (5.5 wt. %) for larger pore widths (27-40 Å) and 20-35 MPa. But in those conditions the volumetric capacity lies between 0.021 and 0.032 kg/L, lower than the DOE 2025 volumetric target.

The ultimate goal is to find an absorbent able to work at rather low loading pressures, for the reasons indicated in the Introduction of this work. The calculations of the usable capacities presented above were carried out for a depletion pressure of 0.5 MPa. The loading-depletion pressure window can be easily modified to scan different sorption-desorption conditions. In Fig. 12, the usable capacities at room temperature have been plotted as a function of the loading pressure for two different values of the depletion pressure, 0.1 MPa (room pressure) and 0.5 MPa. It can be noticed in Fig. 12 that the usable volumetric capacity reaches the DOE 2025 target at a loading pressure of 6.6 MPa when the depletion pressure is 0.1 MPa. This is a quite relevant result of the present work. To the best of our knowledge, no other absorbent material with usable volumetric capacity beyond the DOE 2025 target, and working at such a low loading pressure, has been proposed so far. Such a low loading pressure means that a conformable, cheap, and light tank could be employed; that a much less energy cost would be required to pressurize the gas as compared with current tanks working at 70 MPa (or even at 35 MPa); and that the device would improve in security. The usable gravimetric capacities, on the contrary, do not reach the DOE 2025 target at those conditions.

With a more general perspective, we want to emphasize that our methodology allows to determine the usable hydrogen storage capacities in porous materials characterized by a pore size distribution, which is the common scenario. In that case, the hydrogen storage capacities could be easily calculated as the average of the data obtained for the different pore widths.

## 6. Conclusions

In this work, we have presented an extensive computational study of the hydrogen storage capacities of slit pores constituted by two parallel $BC_3$ sheets decorated with Li atoms. The methodology employed combines density-functional theory, which is used to obtain the structures and the interaction potential energy of a H$_2$ molecule within the Li-decorated BC$_3$ slit pores, and a quantum-thermodynamic model that allows to obtain the adsorbed, total and usable volumetric and gravimetric hydrogen storage capacities. Our results highlight the important role played by the rotational degree of freedom of the H$_2$ molecule in determining the confining potential within the slip pores and their hydrogen storage capacities.

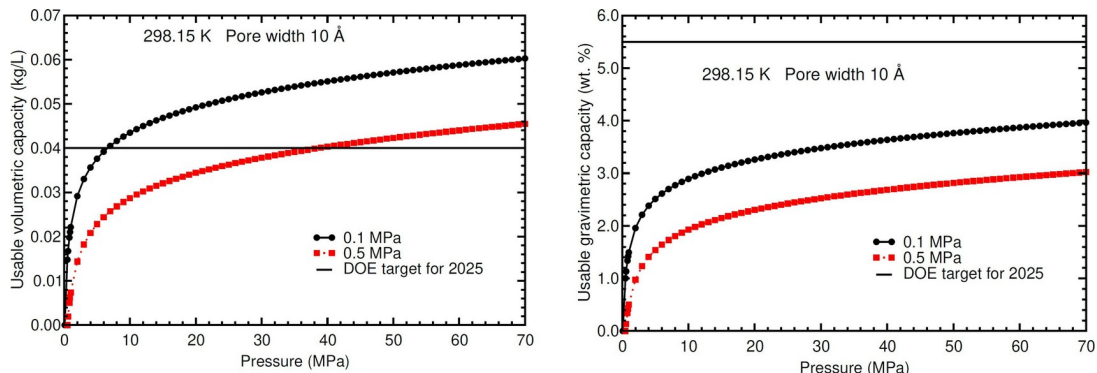

**Fig. 12.** Usable volumetric and gravimetric hydrogen storage capacities of a Li-decorated BC$_3$ slit pore of width 10 Å at 298.15 K, as functions of the loading pressure and the depletion pressure.

Earlier DFT studies on the hydrogen storage capacities of the BC$_3$ sheet and other adsorbents [55–58] were not realistic and overestimated the capacities, because they did not consider pores formed by two parallel sheets, but just one sheet, i.e., they did not investigate the dependence of the hydrogen storage capacity on the pore width, the temperature and the pressure. But the most important flaw of those studies, in our opinion, is that they did not calculate the usable hydrogen storage capacities, which are the relevant quantities when identifying new materials as potential hydrogen storage media, as it has been established by the U.S. DOE.

Our results show that the usable volumetric hydrogen storage capacities of Li-decorated BC$_3$ slit pores of width 9.0-12.0 Å reach the DOE 2025 volumetric target, 0.040 kg/L, at room temperatures, 300-320 K, and moderate pressures, 30-35 MPa, but the usable gravimetric capacities are about two times smaller than the DOE 2025 target, 5.5 wt.%, for those pore widths, temperatures and pressures. Hence, a hydrogen vehicle using an onboard deposit with this porous material inside would have the same autonomy than a fossil fuel based vehicle at room temperatures and moderate pressures, but the deposit would be about two times heavier. We have found that the usable volumetric capacity reaches the DOE 2025 target at a loading pressure of 6.6 MPa when the depletion pressure is 0.1 MPa. This is a quite relevant result of the present work. To the best of our knowledge, no other absorbent material with usable volumetric capacity beyond the DOE 2025 target, and working at such a low loading pressure, has been proposed so far. Such a low loading pressure means that a conformable, cheap, and light tank could be employed, that a much less energy cost would be required to pressurize the gas as compared with current tanks working at high pressures, and that the device would improve in security.

As it is well known, the identification of adsorbent materials meeting all the targets established by the U.S. DOE is a challenging problem. The results presented in this work can be useful for both theoretical and experimental researchers in the area. Real BC$_3$ nanoporous materials should consist in a network of slit-shaped BC$_3$ pores with a pore size distribution in the range of several to hundreds Å. The results of our investigation, in which we have studied pores widths up to 40 Å, strongly suggest that the usable volumetric and gravimetric hydrogen storage capacities of such nanoporous materials, doped with Li atoms, could be probably close to the DOE 2025 targets.

**Acknowledgments**

This research was financially supported by the Spanish MICINN (Grant PGC2018-093745-B-I00) and the Xunta de Galicia (GRC ED431C 2020/10). We also acknowledge the use of the high performance computing equipment of the Pole de Calcul Intensif pour la Mer (DATARMOR, Brest) and the Centro de Proceso de Datos - Parque Cientifico (UVa).

# Supporting Information
## Li-decorated BC$_3$ nanopores. Promising materials for hydrogen storage

I. Cabria[a], A. Lebon[b], M. B. Torres[c], L. J. Gallego[d], A. Vega[a]

a Departamentode Física Teórica, Atómica y Óptica, Universidad de Valladolid, ES-47011 Valladolid, Spain
b Laboratoire de Chimie Electrochimie Moléculaire et Chimie Analytique, Université de Brest, UMR CNRS 6521, F-29285 Brest,France
c Departamento de Matemáticas y Computación, Escuela Politécnica Superior, Universidad de Burgos, ES-09006 Burgos, Spain
d Área de Física de la Materia Condensada, Departamento de Física de Partículas, Facultad de Física, Universidad de Santiago de Compostela, ES-15782 Santiago de Compostela, Spain

1. Local minima of the interaction potential and comparison of the potentials obtained with the different exchange-correlation functional

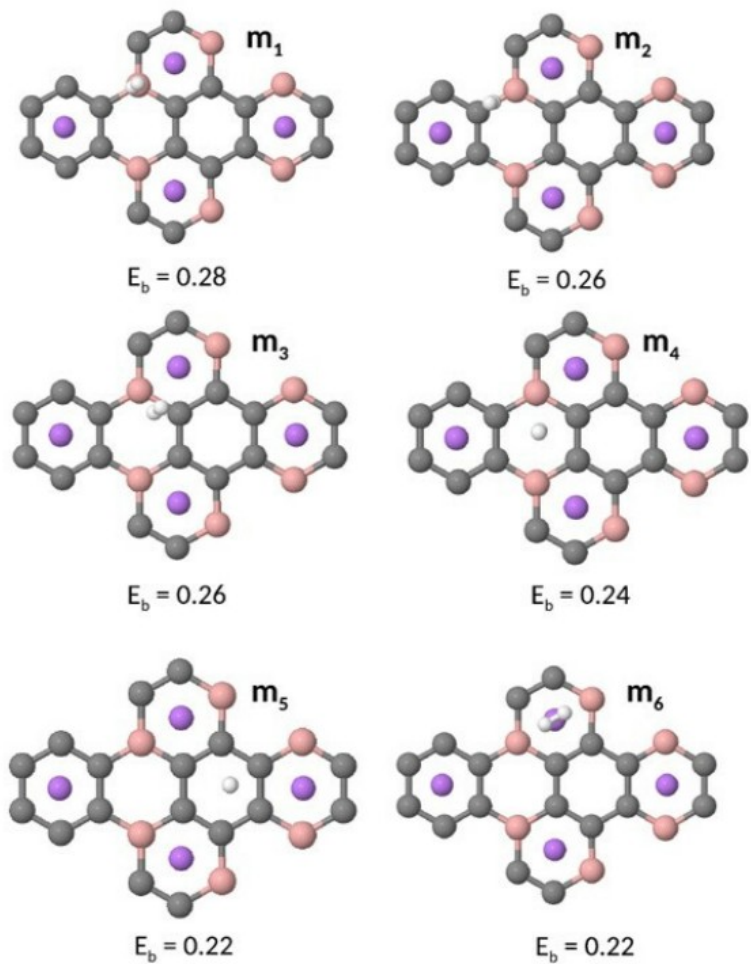

Fig. S1: System composed by the Li-decorated $BC_3$ sheet and the $H_2$ molecule at the six main local minima of the interaction potential. We show the values of $H_2$-binding energies (i.e., the sign-reversed potential energies) in eV.

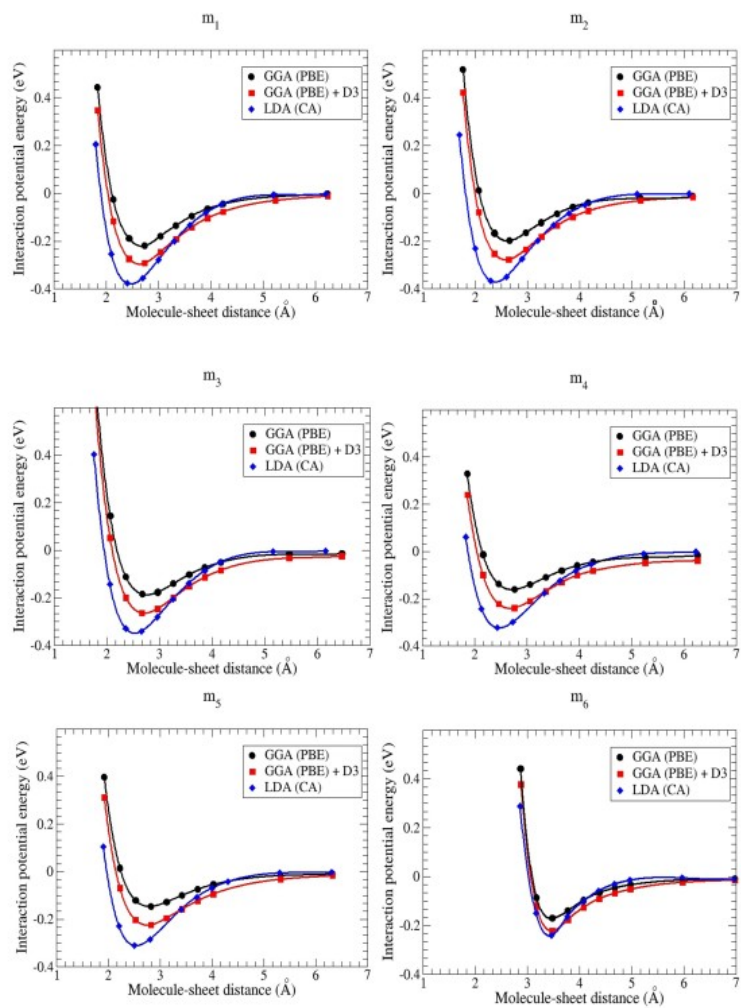

Fig. S2: Interaction potential curves around the six main local minima, as obtained with the three different exchange-correlation functionals.

## 2. Exact and approximated relationships between the adsorbed volumetric and gravimetric hydrogen storage capacities

There is an exact relationship between the adsorbed volumetric and gravimetric hydrogen storage capacities. According to Eq. 6 of the main text, the mass of the adsorbed hydrogen is given by

mass$_{H\ adsorbed}$ (P,T) = $v_c$ (P, T)$V_{pore}$ . In the case of slit-shaped pores, mass $_{adsorbent\ material}$ = $\sigma_{ads}$ 2S , here $\sigma_{ads}$ is the mass surface density of a sheet of the slit pore and S is the surface of one sheet. The origin of the factor 2 is that the slitpore is formed by two sheets. Hence, Eq. 8 of the main text can be written exactly as

$$g_c = \frac{100 v_c \cdot w}{v_c \cdot w + 2\sigma_{ads}} = \frac{100 x}{x+1} \quad \text{(S1)}$$

being x = $v_c$ w/2$\sigma_{ads}$ > 0, a dimensionless magnitude.

We now note that the adsorbed volumetric capacity as a function of the pore width w has some maximum and then decreases slowly towards a constant value (see Fig. 7 of the main text). Hence, according to the exact Eq. S1, the adsorbed gravimetric capacity is an increasing function of the variable x, but not an increasing function of w. To be more precise, $g_c$ is an increasing function of the product $v_c$w, since the mass surface density, $\sigma_{ads}$ , is constant for slit pores of the same type, as can be noticed in Table S1.

Table S1: The mass surface densities, $\sigma_{ads}$, of some sheets in $10^{-26}$ kg/Å$^2$.

| Sheet | $\sigma_{ads}$ |
|---|---|
| Graphene | 0.76 |
| Li-decorated BC$_3$ | 0.73 |

The ranges of values of the adsorbed volumetric capacity, the pore width and the product of these two quantities are shown in Table S2.

Table S2: Ranges of values of the adsorbed volumetric capacity $v_c$ at 77-298.15 K and 0.1-25 MPa, in kg/L, of the pore width $w$, in Å, of the product $v_c w$, in $10^{-26}$ kg/Å$^2$, and of the dimensionless magnitude $x = v_c w/2\sigma_{ads}$, for different types of slit pores.

| $v_c$ | $w$ | $v_c w$ | $x$ |
|---|---|---|---|
| 0.001-0.08 | 5-100 | 0.0005-0.8 | 0.0002-0.4 |

The data in Tables S1 and S2 prove that v c w is smaller or much smaller than 2σ ads . The dimensionless magnitude x in Table S2 shows how much small is v c w compared to 2σ ads : it is between 0.0002 and 0.4. Therefore, it is a reasonable approximation to make x/(x + 1) ≈ x for graphene and Li-decorated BC$_3$ slit pores with a small value of x. Applying this approach to Eq. S1, we obtain the approximate equation

$$g_c \approx 100\,x = 100\,\frac{v_c \cdot w}{2\sigma_{ads}} \quad \text{(S2)}$$

The exact and the approximated (Eq. S2) adsorbed gravimetric capacities for Li-decorated $BC_3$ slit pores at 298.15 K and 25 MPa are compared in Fig. S3. As it can be noticed, the approximation is very good.

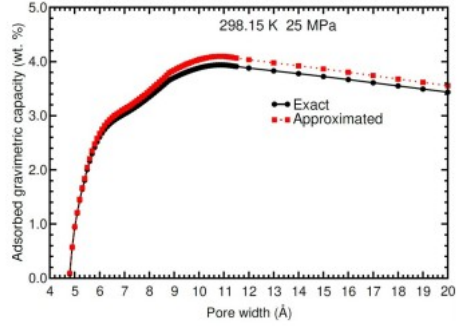

Fig. S3: Exact and approximated (Eq. S2) adsorbed gravimetric capacities of Li-decorated $BC_3$ slit pores as functions of the pore width at 298.15 K and 25 MPa.

When comparing the adsorbed gravimetric capacities of graphene and Li-decorated $BC_3$ slit pores, it must be taken into account that the surface densities of their corresponding sheets are very similar (see Table S1). Hence, the differences in the adsorbed gravimetric capacities of the slit pores can be mainly ascribed to differences in the adsorbed volumetric capacities. Inserting the approximation $\sigma_{graphene} \approx \sigma_{Li-decorated\ BC_3}$ into Eq. S2, it can be obtained, for the same pore width w, the approximated equation

$$\frac{g_c(BC_3Li)}{g_c(graphene)} \approx \frac{v_c(BC_3Li)}{v_c(graphene)} \quad \text{(S3)}$$

In order to test this equation, we have plotted in Fig. S4 the ratio of the adsorbed gravimetric capacities, $\frac{g_c(BC_3Li)}{g_c(graphene)}$ vs the ratio of adsorbed volumetric capacities, $\frac{v_c(BC_3Li)}{v_c(graphene)}$, at 298.15 K and 25 MPa for pore widths between 4.6 and 20 Å. The graph proves that Eq. S3 is very well satisfied. That equation contains two approximations explained before: $x/(x+1) \approx x$, where $x = v_c w/2\sigma_{ads}$, and $\sigma_{graphene} \approx \sigma_{Li-decoratedBC_3}$.

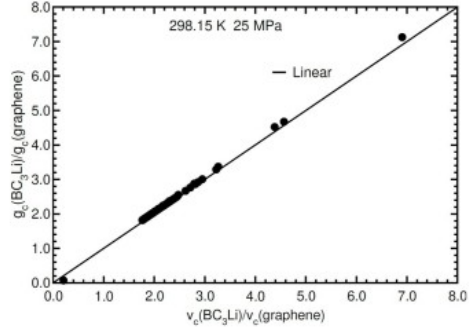

Fig. S4: Comparison of the ratio of the adsorbed gravimetric capacities with the ratio of the adsorbed volumetric capacities of graphene and Li-decorated $BC_3$ slit pores at 298.15 K and 25 MPa for pore widths between 4 and 20 Å.

## 3. Location of highest gravimetric capacity

According to Eq. S2, the adsorbed gravimetric capacity is approximately proportional to the product $v_c w$. Therefore, the maximum value of $g_c$ is not located at the maximum value of $v_c$, but at the maximum value of $v_c w$. Let's call $w_{maxgc}$ the value of the pore width for which the maximum of $g_c$ is found. The derivative of $g_c$ with respect to w must be zero at that point. This implies that

$$\left(\frac{dv_c}{dw}\right)_{w=wmaxgc} w_{maxgc} + v_c\left(w_{maxgc}\right) = 0 \quad (S4)$$

This equation turns into

$$w_{maxgc} = \frac{-v_c\left(w_{maxgc}\right)}{\left(\dfrac{dv_c}{dw}\right)_{w=wmaxgc}} \quad (S5)$$

Now, we focus in the region in which $w > w_{maxvc}$, where $w_{maxvc}$ is the location of the maximum of the volumetric capacity. The derivative of $v_c$ with respect to w is negative in that region (see Fig. 7 of the main text). Hence, Eq. S5 turns finally into

$$w_{maxgc} = \frac{v_c\left(w_{maxgc}\right)}{\left|\left(\dfrac{dv_c}{dw}\right)_{w=wmaxgc}\right|} \quad (S6)$$

The absolute value of the derivative or slope of the adsorbed $v_c$ in that region when the orientation of the $H_2$ molecule is not fixed is smaller than for the fixed orientation. Besides, in the mentioned region, $v_c$ is larger for the non-fixed orientation than for the fixed orientation. Therefore, according to Eq. S6, the pore width corresponding to the maximum of the adsorbed gravimetric capacity ($w_{maxgc}$) for the non-fixed orientation is larger than for the fixed orientation, i.e., the highest adsorbed gravimetric capacities for the non-fixed orientation are located at larger pore widths, as it can be observed in the results shown in Fig. 7 and Table 1 of the main text.

# 4. Additional results for the total and usable hydrogen storage capacities of Li-decorated BC$_3$ slitpores

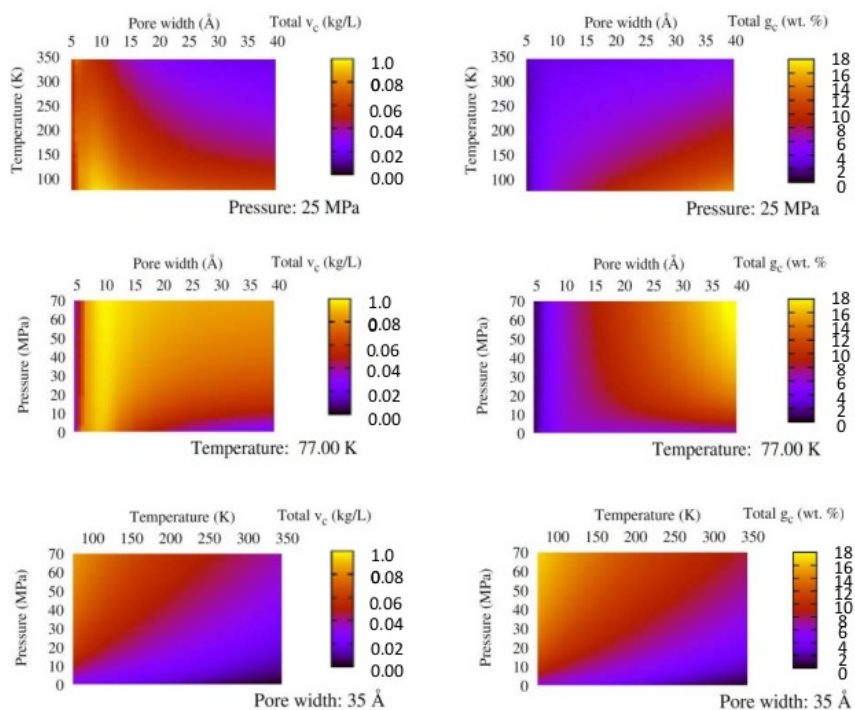

Fig. S5: Total volumetric and gravimetric hydrogen storage capacities of Li-decorated BC$_3$ slit pores as functions of the pore width and temperature at 25 MPa (upper panels), of the pore width and pressure at 77 K (middle panels), and of the temperature and pressure for a pore width of 35 Å (lower panels).

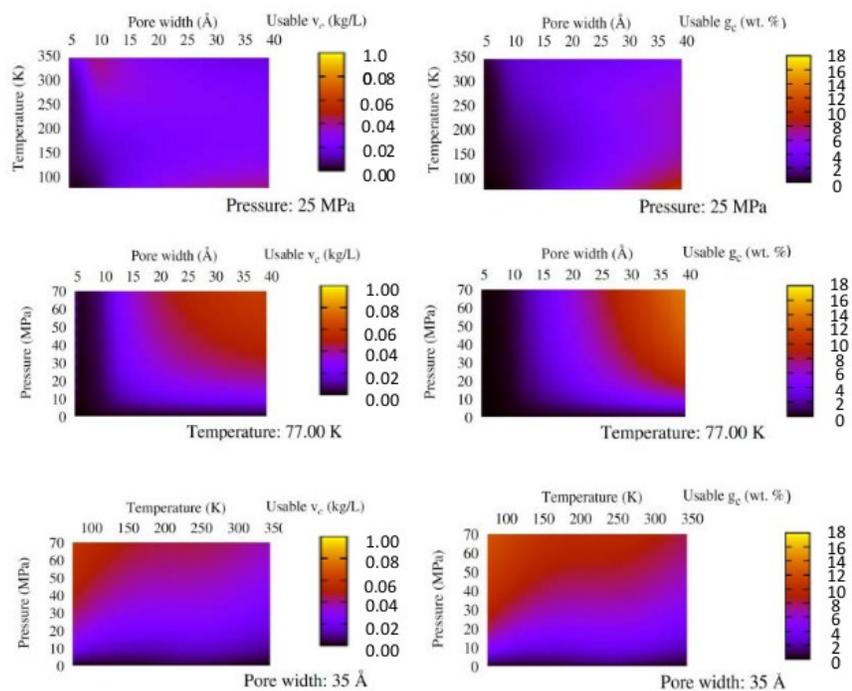

Fig. S6: Usable volumetric and gravimetric hydrogen storage capacities of Li-decorated $BC_3$ slit pores as functions of the pore width and temperature for 25 MPa (upper panels), of the pore width and pressure for 77 K (middle panels) and of the temperature and pressure for a pore width of 35 Å (lower panels).

Acknowledgments This research was financially supported by the Spanish MICINN (Grant PGC2018-093745-B-I00) and the Xunta de Galicia (GRC ED431C 2020/10). We also acknowledge the use of the high performance computing equipment of the Pole de Calcul Intensif pour la Mer (DATARMOR, Brest) and the Centro de Proceso de Datos - Parque Cientifico (UVa).